\documentclass[aps,pre,showpacs,showkeys,twocolumn,preprintnumbers,floatfix,nofootinbib,10pt]{revtex4-1}
\usepackage{amssymb}
\usepackage{graphicx,color}
\usepackage{amscd,amsmath,verbatim}
\usepackage{amsfonts,epsfig}
\usepackage{mathptmx}
\usepackage{setspace}
\usepackage{epsfig}
\usepackage{url}
\usepackage{multirow}
\usepackage{color}
\usepackage{subfigure}

\begin{document}

\title{Unveiling the ZGB model with $CO$ desorption: a single model with two
universality classes?}
\author{Henrique A. Fernandes$^{1}$, Roberto da Silva$^{2}$, and Alinne B.
Bernardi$^1$}

\affiliation{$^1$Instituto de Ci{\^e}ncias Exatas, Universidade Federal de Goi{\'a}s, Regional Jata{\'i}, BR 364, km 192, 3800 - CEP 75801-615, Jata{\'i}, Goi{\'a}s, Brazil \\
$^2$Instituto de F{\'i}sica, Universidade Federal do Rio Grande do Sul, Av. Bento Gon{\c{c}}alves, 9500 - CEP 91501-970, Porto Alegre, Rio Grande do Sul, Brazil}

\begin{abstract}
We study the behavior of the phase transitions of the Ziff-Gullari-Barshad
(ZGB) model when the $CO$ molecules are adsorbed on the catalytic surface
with a rate $y$ and desorbed from the surface with a rate $k$. We employ
large-scale nonequilibrium Monte Carlo simulations along with an
optimization technique based on the coefficient of determination, in order
to obtain an overview of the phase transitions of the model in the whole
spectrum of $y$ and $k$: ($0\leq y\leq 1$ and $0\leq k\leq 1$) with
precision $\Delta y=\Delta k=0.001$. Sucessive refinements reveal a region
of points belonging to the directed percolation universality class whereas
the exponents $\theta $ and $\beta /\nu _{\parallel }$ obtained agree with
those of this universality class. On the other hand, the effects of allowing
the $CO$ desorption from the lattice on the discontinuous phase transition
point of the original ZGB model suggest the emergence of an Ising-like point
previously predicted in Ref. \cite{tome1993}. We show that such a point
appears after a sequence of two lines of pseudo-critical points which leads
to a unique peak of the coefficient of determination curve in $y_{c}=0.554$
and $k_{c}=0.064$. In this point, the exponent $\theta $ agrees with the
value found for Ising model.
\end{abstract}

\maketitle

\section{Introduction}

\label{sec:introduction}

Over several decades, the study of the critical behavior of many-body
systems has been mainly carried out through Monte Carlo simulations which
makes it one of the most important methods in statistical mechanics. At the
very beginning, to circumvent the critical slowing down (characteristic of
the long-time regime) of systems close to their critical point was not a
simple task. However, nowadays, with the advances in computational
technology and with the discovery of new techniques, this is not anymore the
main concern when studying phase transitions and critical phenomena of those
systems. In 1989, Janssen, Schaub, and Schmittmann \cite{janssen1989}, and
Huse \cite{huse1989} proposed a method which avoids the critical slowing
down and is known as short-time (nonequilibrium) Monte Carlo simulations.
They discovered, by using renormalization group techniques and numerical
calculations, respectively, that there is universality and scaling behavior
even at the early stage of the time evolution of dynamical systems \cite%
{zheng1998}. Since then, this method has been successfully applied to a wide
variety of problems ranging from systems with defined Hamiltonian \cite%
{zheng2001, albano2001a, silva2002a, silva2002b, arashiro2003, silva2004a,
grandi2004, fernandes2005, hadjiagapiou2005, fernandes2006a, fernandes2006b,
fernandes2006c, prudnikov2010, silva2013a, silva2013b, silva2014,
chiocchetta2016} to models based on generalized Tsallis statistics \cite%
{silva2012}, protein folding models \cite{arashiro2007}, and models without
defined Hamiltonian such as polymers \cite{luo2006}, contact process and
cellular automaton \cite{silva2004b}, epidemic models \cite{silva2015},
driven lattice gases \cite{basu2017}, model of liquids \cite%
{loscar2016,loscar2017}, and even for surface reaction models \cite%
{albano2001b, fernandes2016}. (An interesting review on the progress of this
method was published by Albano \textit{et. al} \cite{albano2011}.)

The surface reaction models \cite{evans1991a, albano1994, andrade2010,
andrade2012} have attracted considerable interest whereas they possess phase
transitions and critical phenomena and, in addition, can be used to explain
several experimental observations in catalysis \cite{ehsasi1989,
christmann1991, imbhil1995}. One such model was proposed in 1986 by Ziff,
Gulari, and Barshad \cite{ziff1986} to describe some nonequilibrium aspects
of the catalytic reaction of carbon monoxide $CO$ and oxygen $O$ on a
surface to produce carbon dioxide $CO_2$ ($CO + O \longrightarrow CO_2$). In
this model, also known as ZGB model, the surface is represented by a
two-dimensional regular square lattice. Both $CO$ and $O_2$ molecules in the
gas phase impinge the surface at rates $y$ and $1-y$, respectively. Each $CO$
molecule which impinges the surface needs only one vacant site to be
adsorbed on it. However, the $O_2$ molecule dissociates into two $O$ atoms
during its adsorption process. The atoms are both adsorbed on the surface
when there exists two vacant nearest-neighbor sites. The production of $CO_2$
molecules occur when, after the adsorption processes, a nearest-neighbor
pair of $CO$ and $O$ is found. In this case, the $CO_2$ molecule desorbs and
returns to the gas phase, leaving the surface with two vacant sites. This
model possesses only one control parameter, the $CO$ adsorption rate $y$
(the partial pressure of $CO$ in the gas phase), and exhibits three distinct
states and two irreversible phase transitions: one continuous and another
discontinuous. The continuous phase transition occurs at $y=y_1 \sim 0.3874$ 
\cite{voigt1997} and separates the $O$ poisoned state ($0 \leq y < y_1$)
from the reactive state ($y_1 < y < y_2$) where both $CO$, $O$, and vacant
sites coexist on the catalytic surface, and there is sustainable production
of $CO_2$ molecules. The discontinuous phase transition occurs at $y=y_2
\sim 0.5256$ \cite{ziff1992} and separates the reactive state from the $CO$
poisoned state ($y_2 < y \leq 1$). So, despite the simplicity of this model,
its rich phase diagram, experimental observations, and industrial
applications have made the ZGB model one of the most proeminent examples in
the study of reaction processes on catalytic surfaces \cite{meakin1987,
dickman1986,fischer1989,marro1999}. Several modified versions of the model
have been proposed in order to obtain more realistic systems of actual
catalytic processes, for instance, by including $CO$ desorption \cite%
{tome1993,fischer1989,dumont1990, albano1992, kaukonen1989, jensen1990,
brosilow1992,matsushima1979, buendia2013, buendia2015, chan2015}, diffusion 
\cite{ehsasi1989, fischer1989, kaukonen1989, jensen1990, grandi2002,
buendia2015}, impurities \cite{hoenicke2000, buendia2012, buendia2013,
buendia2015, hoenicke2014}, attractive and repulsive interactions between
the adsorbed molecules \cite{buendia2015, satulovsky1992}, surfaces of
different geometries \cite{meakin1987, albano1990}, and with hard oxygen
boundary conditions \cite{brosilow1993}, etc.

The ZGB model has been studied through several techniques.\textbf{\ }One of
them was considered recently by the authors and will be also employed in
this work to study the ZGB model with desorption of $CO$ molecules. In that
work \cite{fernandes2016}, we studied the original ZGB model through
short-time Monte Carlo simulations and considered a method which is based on
optimization of the coefficient of determination of the order parameter, in
order to locate its nonequilibrium phase transitions. This amount, which
measures the quality of a linear fit, is widely employed in Statistics and
was proposed for the first time in Statistical Mechanics by one of the
authors of this current work and other collaborators to optimize the
critical temperature of spin systems \cite{silva2012}. Since then, this
method has been successfully applied in other systems with and without
defined Hamiltonian (see, for example: \cite{silva2013a,
silva2013b,silva2014,silva2015,Fernandes2017})

As shown in Ref. \cite{fernandes2016}, this method was able to characterize
the the continuous phase transition of the original ZGB model and its upper
spinodal point, as well as estimate the static and dynamic critical
exponents which are in complete agreement with results found in literature.

By taking into consideration those unambiguous results obtained through
short-time Monte Carlo simulations along with the coefficient of
determination, we decided to study a modified version of the ZGB model to
include the desorption of $CO$ molecules \cite{kaukonen1989, brosilow1992,
albano1992, tome1993} from the catalytic surface in order to obtain a
detailed framework of the phase diagram of the model. We do not consider the
desorption of $O$ molecules whereas one has been shown that the desorption
rate $k$ of $CO$ molecules is much higher than that of $O$ atoms \cite%
{ehsasi1989}. Physically, the desorption of $CO$ molecules can be thought of
as the temperature effect which is a very important parameter in catalytic
processes. In addition, the desorption of $CO$ molecules from the surface
also prevents the appearance of the $CO$ poisoned phase, which turns the
discontinuous phase transition reversible \cite{matsushima1979, buendia2013}%
. Although thinking of the desorption of $CO$ molecules as an effect of
temperature, in this work we do not include other correlated effects such as
the diffusion of $CO$ molecules and/or $O$ atoms on the surface since we are
only concerned with the analysis of the ZGB model with the desorption
process. So, this model has now two control parameters, the $CO$ adsorption
rate $y$ and the $CO$ desorption rate $k$. Our results show a unprecedented
coexistence of two universality classes in the same model: the directed
percolation (DP) universality class in the region of the continuous phase
transition of the original model, as well as a single point that apparently
has Ising-like characteristics as suggested by T. Tom\'{e} and R. Dickman 
\cite{tome1993}.

The rest of this paper is organized as follows. In Sec. II, we present the
modified version of the ZGB model to include the desorption of $CO$
molecules from the surface and describe the short-time Monte Carlo
simulations as well as the technique known as coefficient of determination
(see, for example, Ref. \cite{trivedi2002}). In Sec. III, we present our
main results by showing how these techniques can be used to obtain a
framework of the phase diagram of the model. In that section, as additional
results, we also estimate some critical exponents for some specific critical
points ($y_{c},k_{c}$). Finally, a brief summary is given in Sec. IV.

\section{Model and simulation method}

\label{sec:model}

The ZGB model \cite{ziff1986} simulates the catalytic oxidation obtained
with the reaction between carbon monoxide ($CO$) molecules and oxigen ($O$)
atoms on a surface which, in turn, is in contact with a gas phase composed
by $CO$ and $O_2$ molecules.

This catalytic surface can be modelled as a regular square lattice and its
sites might be occupied by $CO$ molecules or by $O$ atoms or may be vacant ($%
V$). The reactions presented in the previous section follow the
Langmuir-Hinshelwood mechanism \cite{ziff1986, evans1991b} and are
schematically represented by the following reaction equations: 
\begin{equation}
CO(g)+V\longrightarrow CO(a)  \label{eq:co_ad}
\end{equation}%
\begin{equation}
O_{2}(g)+2V\longrightarrow 2O(a)  \label{eq:o2_ad}
\end{equation}%
\begin{equation}
CO(a)+O(a)\longrightarrow CO_{2}(g)+2V  \label{eq:co2_des}
\end{equation}%
where $g$ and $a$ refer, respectively, to the gas and adsorbed phases of the
atoms and molecules. The Eq. (\ref{eq:co_ad}) takes into account the
adsorption process of $CO$ molecules on the surface, i.e., if a $CO$
molecule is selected in the gas phase (with a rate $y$), a site on the
surface is chosen at random and, if it is vacant $V$, the molecule is
immediately adsorbed at this site. Otherwise, if the chosen site is
occupied, the $CO$ molecule returns to the gas phase and the trial ends. The
Eq. (\ref{eq:o2_ad}) considers the adsorption process of $O_{2}$ molecules,
i.e., if an $O_{2}$ molecule is selected in the gas phase (with a rate $1-y$%
), then a nearest-neighbor pair of sites is chosen at random. If both sites
are empty, the $O_{2}$ molecule dissociates into a pair of $O$ atoms which
are adsorbed on these sites. However, if one or both sites are occupied, the 
$O_{2}$ molecule returns to the gas phase and the trial ends. Finally, after
each adsorption event, all nearest-neighbor sites of the newly occupied site
are checked randomly. If one $O-CO$ pair is found, they react immediately
forming a $CO_{2}$ molecule which desorbs leaving behind two vacant sites
(Eq. (\ref{eq:co2_des})).

In this work, we modified the ZGB model by including the desorption of $CO$
molecules with a rate $k$ whose equation is given by 
\begin{equation}
CO(a)\longrightarrow CO(g)+V.  \label{eq:co_des}
\end{equation}%
This equation accounts for the possibility, observed in experiments, of the
desorption of $CO$ molecules adsorbed on the catalytic surface without
reacting. Here, we do not consider the desorption of $O$ atoms since, as
shown in Ref. \cite{ehsasi1989}, its desorption rate is much smaller than $k$%
. As pointed out above, the desorption can be thought of as an effect of the
temperature, i.e., the higher the temperature, the bigger the energy of the
molecule and so, the bigger the probability of desorption.

The study is carried out via short-time Monte Carlo (MC) simulations in
order to obtain a framework of the phase diagram of the ZGB model with
desorption of $CO$ molecules. To perform the numerical simulations, we take
into consideration that, for systems with absorbing states, the finite size
scaling near criticality can be described by the following general scaling
relation \cite{hinrichsen2000}: 
\begin{equation}
\left\langle \rho(t)\right\rangle \sim
t^{-\beta/\nu_{\parallel}}f((y-y_c)t^{1/\nu_{\parallel}},t^{d/z}L^{-d},
\rho_0t^{\beta/\nu_{\parallel}+\theta}),  \label{eq:fss}
\end{equation}
where $\rho$ is the density of vacant sites (the order parameter of the
model), $\left\langle \cdots \right\rangle$ stands for the average on
different evolutions of the system, $d$ is the dimension of the system, $L$
is the linear size of a regular square lattice, and $t$ is the time. The
indexes $z=\nu_{\parallel}/\nu_{\perp}$ and $\theta=\frac{d}{z}-\frac{2\beta%
}{\nu_{\parallel}}$ are dynamic critical exponents, and $\beta$, $%
\nu_{\parallel}$, and $\nu_{\perp}$ are static ones. Here, $y-y_c$ is the
distance of a point $y$ to the critical point, $y_c$, which governs the
algebraic behaviors of the two independent correlation lengths: the spatial
one, $\xi_{\perp}\sim (y-y_c)^{-\nu_{\perp}}$, and the temporal one, $%
\xi_{\parallel}\sim (y-y_c)^{-\nu_{\parallel}}$.

The density of vacant sites is given by 
\begin{equation}
\rho(t)=\frac{1}{L^{d}}\sum_{i=1}^{L^{d}} s_{i},
\end{equation}
and $s_i=1$ when the sites $i$ are vacant, otherwise, it is equal to zero.

The critical exponents of the model can be estimated through the Eq. (\ref%
{eq:fss}) and from nonequilibrium Monte Carlo simulations by taking into
account three different initial conditions at criticality. When all sites of
the lattice are initially vacant (with initial density $\rho _{0}=1$), it is
expected that the density of vacant sites decays algebraically as 
\begin{equation}
\left\langle \rho (t)\right\rangle \sim t^{-\beta /\nu _{\parallel
}}=t^{-\delta },  \label{eq:p1}
\end{equation}%
and when the simulation starts with all sites filled with $CO$ molecules ($%
\rho _{0}=0$), or with $O$ atoms, except for a single empty site located in
the center of the lattice, i.e., $\rho _{0}=1/L^{2}$, we expect 
\begin{equation}
\left\langle \rho (t)\right\rangle \sim \rho _{0}t^{\frac{d}{z}-2\frac{\beta 
}{\nu _{\parallel }}}=\rho _{0}t^{\theta }.  \label{eq:p2}
\end{equation}%
So, the exponents $\delta $ and $\theta $ are given by the slope of the
power laws in $\log \times \log $ scale, respectively.

The algebraic behavior presented above along with other power laws found at
the critical point (see, for example, Ref. \cite{silva2004b,fernandes2016}),
both obtained via the short-time MC method, allows us to find the whole set
of critical exponents of the model without the problem of critical slowing
down. So, instead of waiting the system to achieve the steady state to
perform the statistics, which in turn takes between $10^{4}$ to $10^{8}$ MC
steps depending on the lattice size and other parameters, we consider only a
few hundreds of MC steps at the beginning of the time evolution of dynamical
systems to obtain our estimates. In addition, this technique also is used to
estimate the critical points whereas the Eqs. (\ref{eq:p1}) and (\ref{eq:p2}%
) hold only at the criticality. So, out of criticality, those equations are
not straight lines in $\log \times \log $ scale. This is the main reason for
the use of the coefficient of determination to estimate the critical points
of the model. This coefficient is given by 
\begin{equation}
r=\frac{\sum\limits_{t=N_{\min }}^{N_{MC}}(\overline{\ln \left\langle \rho
(t)\right\rangle }-a-b\ln t)^{2}}{\sum\limits_{t=N_{\min }}^{N_{MC}}(%
\overline{\ln \left\langle \rho (t)\right\rangle }-\ln \left\langle \rho
\right\rangle (t))^{2}},  \label{eq:coef_det}
\end{equation}%
where $N_{MC}$ is the number of MC steps, $\rho (t)$ is obtained for each
pair of the control parameters of the model ($y,k$), $a$ is the intercept
and $b$ the slope of a linear function, and $\overline{\ln \left\langle \rho
(t)\right\rangle }=(1/N_{MC})\sum\nolimits_{t=N_{\min }}^{N_{MC}}\ln
\left\langle \rho \right\rangle (t)$, and $N_{\min }$ is the number of MC
steps discarded at the beginning (the first steps). The coefficient $r$ has
a very simple interpretation: it is the ratio (expected variation)/(total
variation) and ranges from 0 to 1. So, the bigger the $r$ ($r\simeq 1$), the
better the linear fit in $\log \times \log $ scale, and therefore, the
better the power law which corresponds to the critical point ($y_{c},k_{c}$%
). On the other hand, when the system is out of criticality, there is no
power law and $r\simeq 0$. Thus, we can obtain the coefficient of
determination for several pairs ($y,k$) by using, for instance, the Eqs. (%
\ref{eq:p1}) or (\ref{eq:p2}).

In this work, we consider the Eq. (\ref{eq:p1}) and obtain the coefficient
of determination $r$ for $10^6$ pairs ($y,k$), i.e., we perform simulations
in the whole spectrum of values of $y$ and $k$ ($0 \leq y \leq 1$ and $0
\leq k \leq 1$ with $\Delta y = \Delta k = 0.001$) in order to have a
framework of the phase diagram of the model, as well as a clue of what
happens with the continuous and discontinuous phase transitions of the
original model when $CO$ desorption is allowed.

\section{Results}

\label{sec:results}

In this section, we present our main results by using large-scale short-time
MC simulations. First, we consider the coefficient of determination to
obtain the framework of the phase diagram of the model. The results allow us
to observe in detail the changes caused by the inclusion of the $CO$
desorption in the original ZGB model by obtain its possible critical points (%
$y_{c},k_{c}$). With these points in hand, we estimate some critical
exponents and compare our results with those ones found in literature.

\subsection{Coefficient of determination}

\begin{figure}[th]
\begin{center}
\includegraphics[width=\columnwidth]{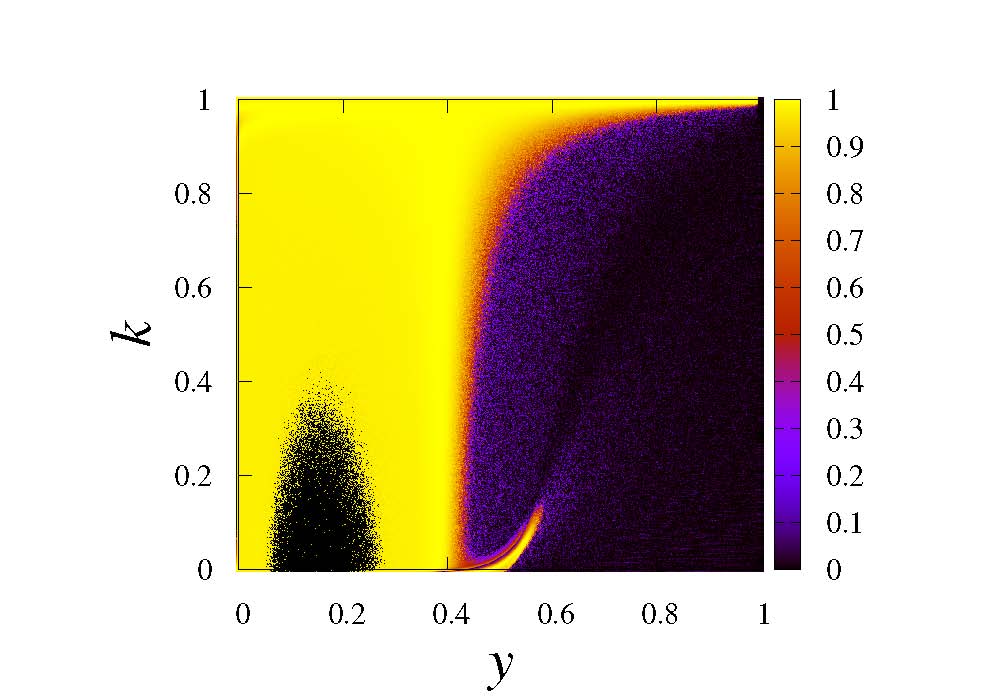}
\end{center}
\caption{Coefficient of determination $r$ as function of $y$ and $k$ for the
ZGB model with desorption of $CO$ molecules.}
\label{fig:coef1}
\end{figure}

Figure \ref{fig:coef1} shows a framework of the phase diagram of the ZGB
model with $CO$ desorption obtained through the coefficient of determination 
$r$ given by Eq. ({\ref{eq:coef_det}}) and by taking into consideration the
Eq. (\ref{eq:p1}). 
To construct this figure, we needed to perform $10^{6}$ independent
simulations. Each simulation was carried out on square lattices of linear
size $L=80$ with periodic boundary conditions and returned the value of $r$
obtained for a given set of the control parameters, $y$ and $k$. Here, we
considered the system starting with a lattice fully ocuppied by $CO$
molecules, which means $\rho _{0}=0$ since we are using the density of empty
sites as the order parameter of the model. Each value of $r$ is an average
taken over 1000 samples in their first 300 MC steps without taking into
account the first $N_{\min }=100$ MCsteps. As stated above, $r$ ranges from
zero (black points) which means that the considered point ($y,k$) is not
critical, to one (yellow points) which means that this point follows a power
law and therefore is a candidate to critical point. This figure shows a
large extension of yellow points, mainly before the critical point of the
original ZGB model, $y\simeq 0.39$ (here named as region 1). As can be seen,
there is also a small region near the discontinuous point, around $%
0.45\lesssim y\lesssim 0.6$ and small values of $k$, which has a set of
yellow points and looks like a tail (the region 2). These behaviors can be
easily explained: as the second order phase transition of the original model
takes place at the point which separates the $O$ poisoned phase from the
beginning of the reactive state where the production of $CO_{2}$ molecules
starts, any value of $k$ does not influence this transition substantially.
However, for the discontinuous phase transition this argument does not hold
since at this point, even a small value of $k$ can avoid the poisoning of
the surface with $CO$ molecules and, therefore, $k$ substantially influences
the transition to the point of eliminating the discontinuous phase
transition.

\begin{figure}[th]
\begin{center}
\includegraphics[width=\columnwidth]{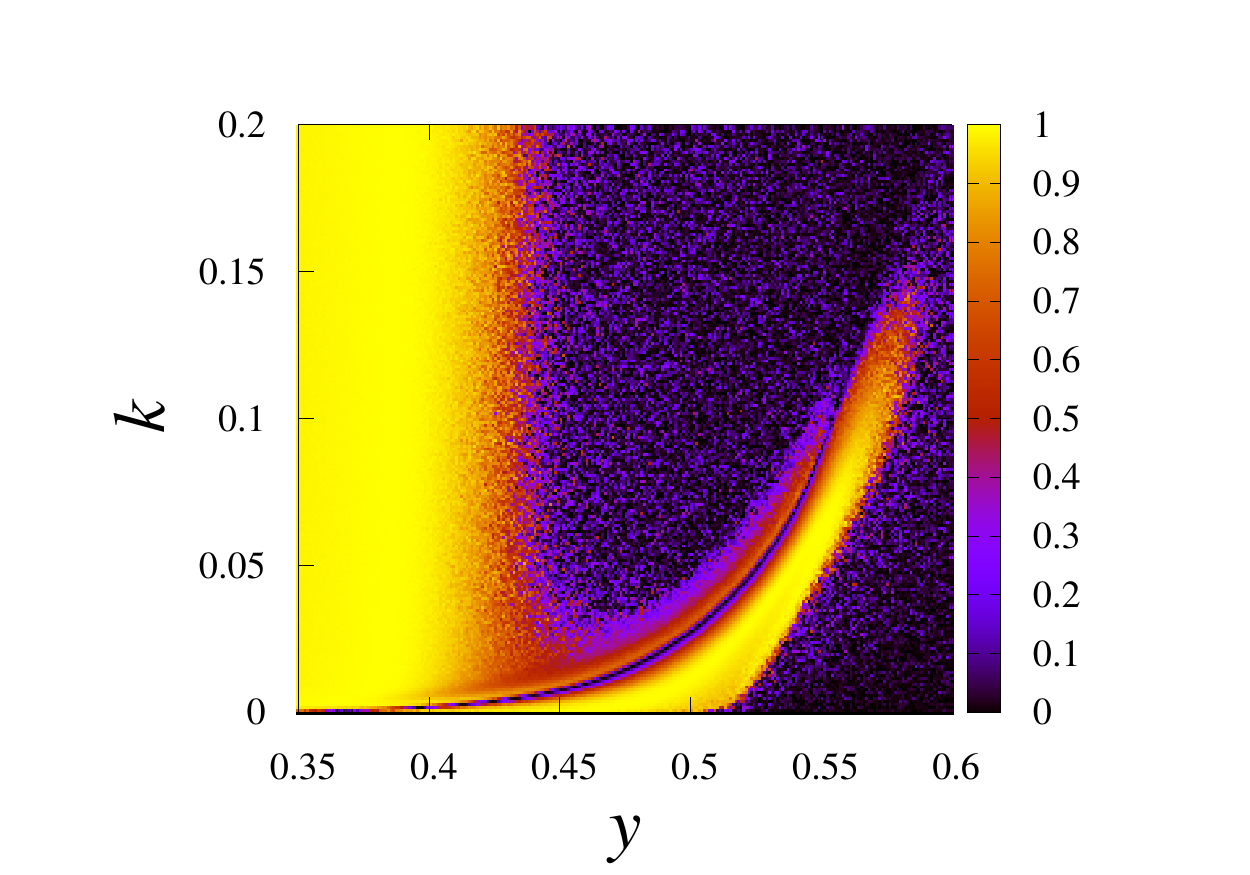}
\end{center}
\caption{A highlight of the coefficient of determination $r$ as function of $%
y$ and $k$ for the ZGB model with desorption of $CO$ molecules. We can
observe an extension that ends after the discontinuous phase thansition
point ($y\approx 0.53$) of the model without desorption ($k=0$). This
extension comes from continuous phase transition of the original ZGB model ($%
y\approx 0.39$ and $k=0$).}
\label{fig:coef2}
\end{figure}

In Fig. \ref{fig:coef2}, we highlight the region that presents an extension
coming from the continuous phase transition ($y\approx 0.39$ and $k=0$) and
that lasts up to $y\approx 0.58$ and $k\approx 0.15$ which is close to
discontinuous phase transition of the original model $y\approx 0.53$ and $k=0
$. \textbf{\ } 

\begin{figure*}[tbh]
\begin{center}
\includegraphics[width=0.85\columnwidth]{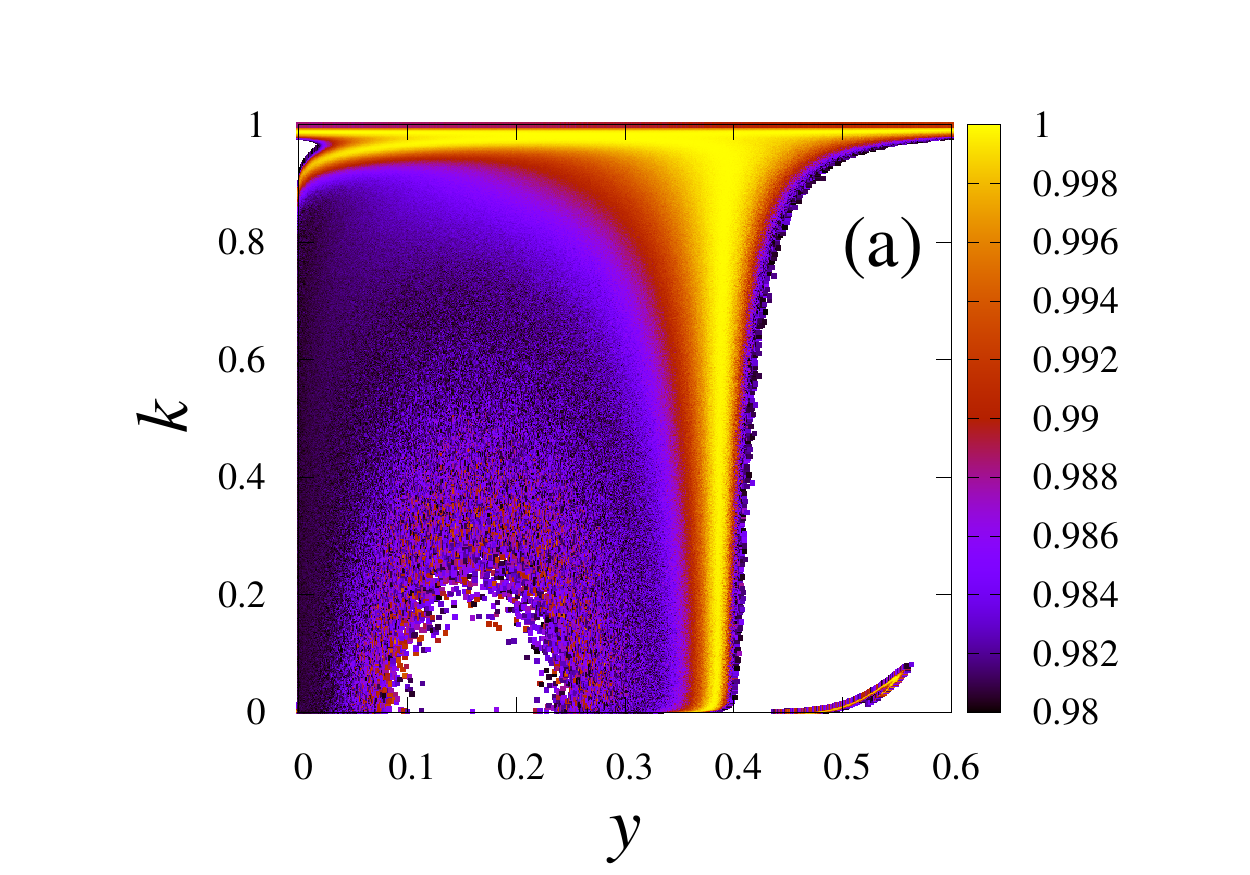}%
\includegraphics[width=0.85\columnwidth]{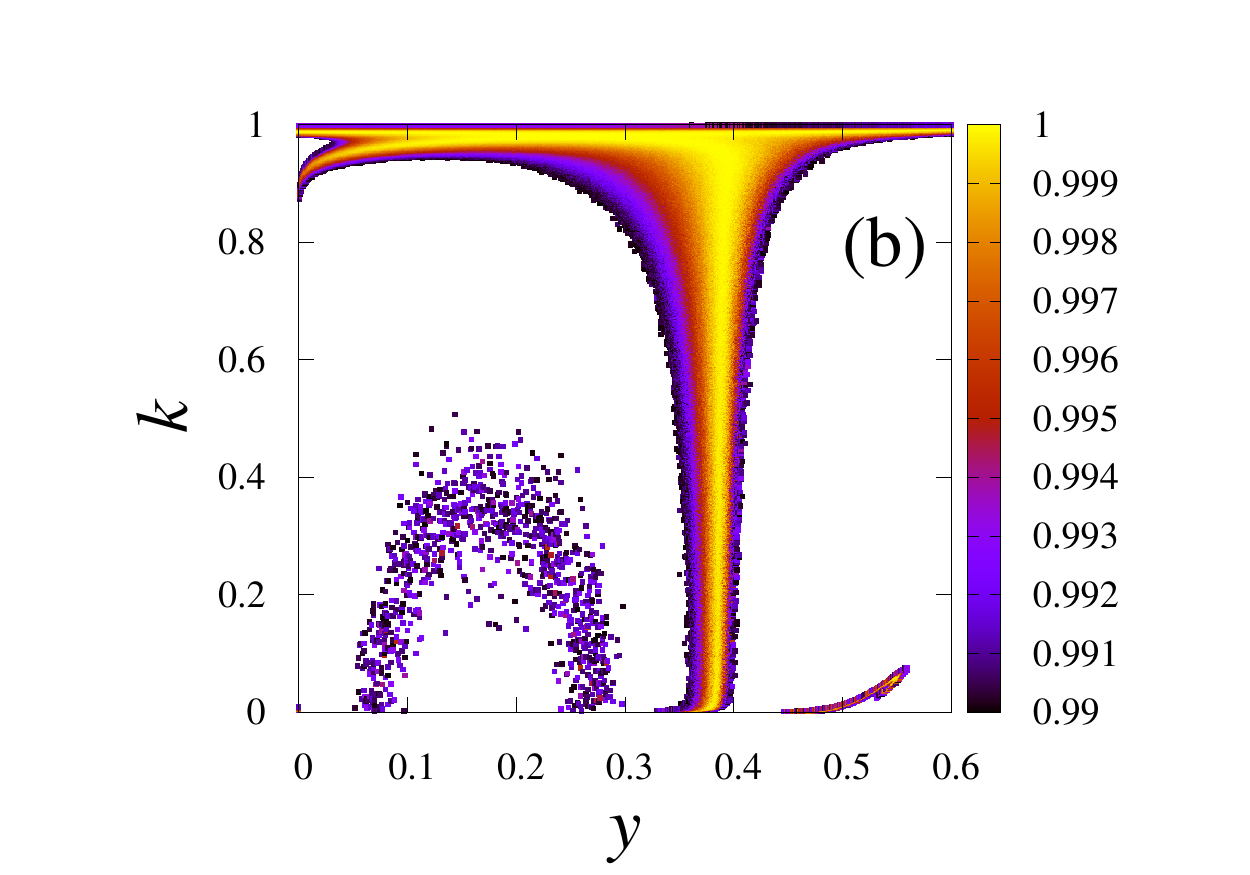} %
\includegraphics[width=0.85\columnwidth]{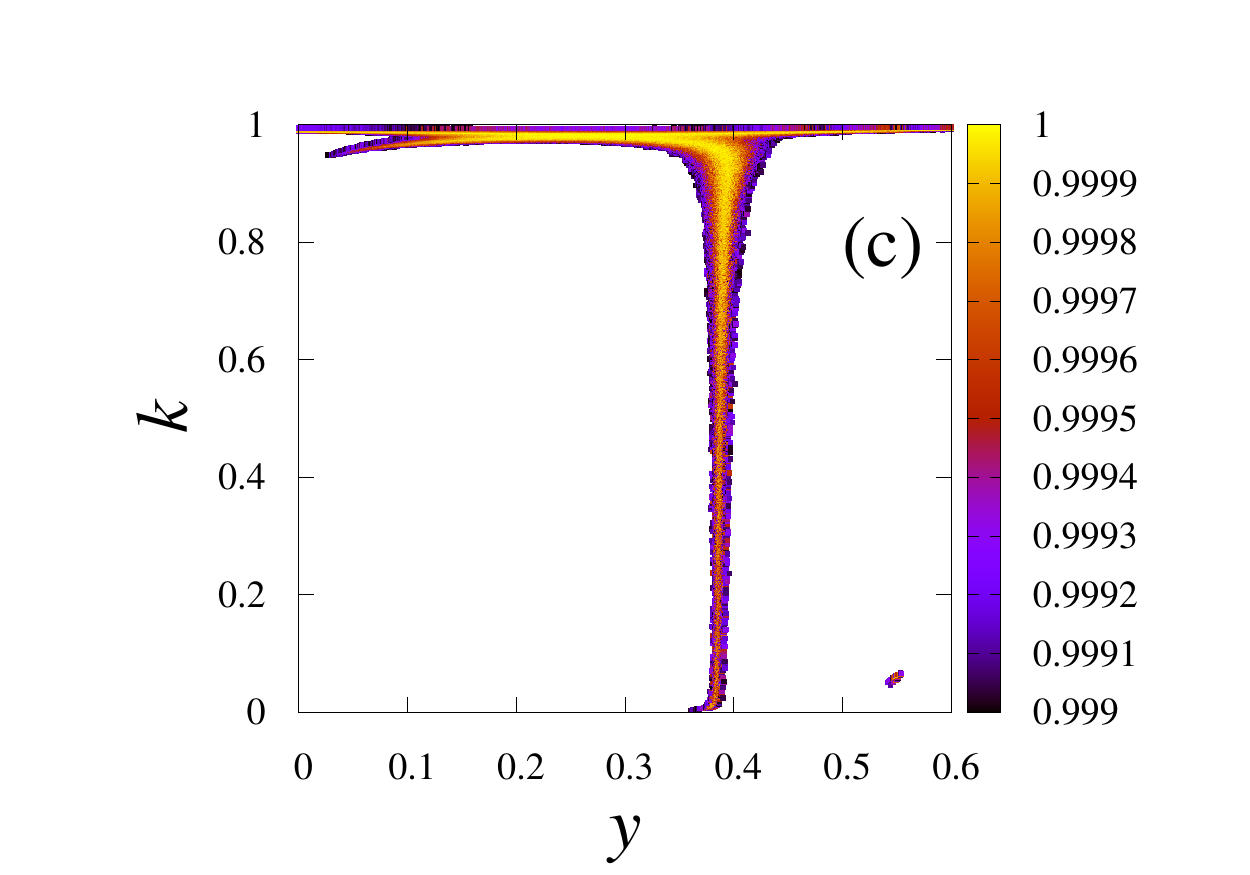}%
\includegraphics[width=0.85\columnwidth]{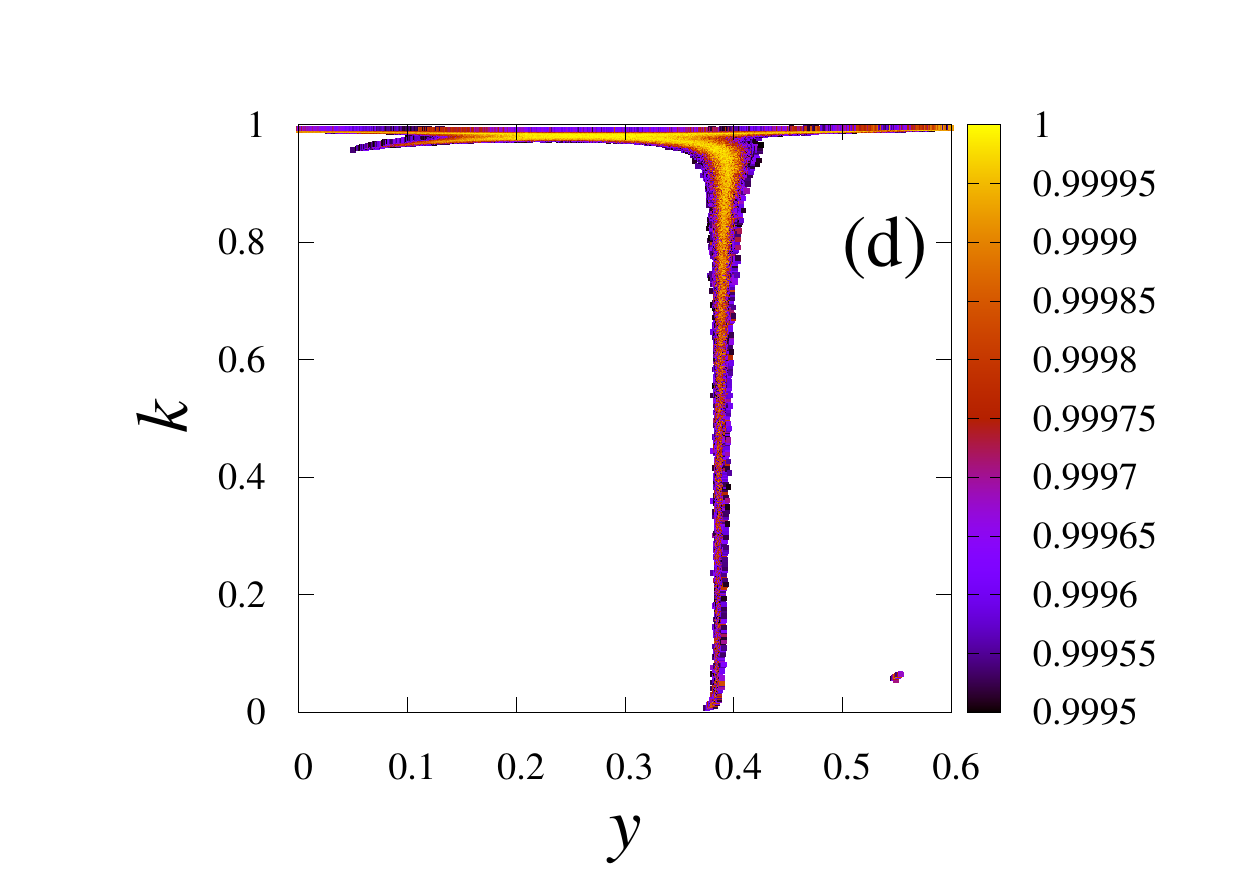}
\end{center}
\caption{Refinement process of the coefficient of determination for (a) $%
r\geq 0.98$, (b) $r\geq 0.99$, (c) $r\geq 0.999$, and (d) $r\geq 0.9995$.}
\label{fig:refinement}
\end{figure*}

Before analysing these two regions in detail, it is important to improve
Fig. \ref{fig:coef1} in order to find points with best coefficients of
determination. Figure \ref{fig:refinement} shows the refinement of Fig. \ref%
{fig:coef1} for higher values of $r$: $r\geq 0.98$, 0.99, 0.999, and 0.9995. 
This figure shows a narrowing of the yellow points in both regions when one
considers only $r\geq 0.98$ to $r\geq 0.9995$. Figure \ref{fig:refinement}
(b) shows that, for $r\geq 0.99$, the region 1 splits into two parts: the
first one looks like an upside-down bow and the second one is a straight
vertical line which starts at $y\simeq 0.39$ and $k=0$ and increases to
around $k=1$ when two branches arise in both sides of the diagram. On the
other hand, the region 2 presents two well defined lines which meet each
other in $y\sim 0.56$ and $k\sim 0.07$. A similar analysis can be done for $%
r\geq 0.999$ (see Fig. \ref{fig:refinement} (c)) with only a decrease of
yellow points. So, as $r\geq 0.999$ is a good value for the coefficient of
determination, all these points can be considered as candidates to phase
transition points. However, as we are looking for higher values of $r$, we
consider in our study only values of $r\simeq 0.9995$ as presented in Fig. %
\ref{fig:refinement} (d). This figure shows that the region 1, around the
critical adsorption rate of the original model ($y\simeq 0.39$) and for all
values of $k$, is not affected by the desorption of $CO$ molecules \cite%
{albano1992, chan2015}. However, the discontinuous phase transition of the
original ZGB model ($y\simeq 0.53$) and $k\simeq 0$ disappears even for very
small values of $k$ \cite{tome1993,chan2015}. In this case, only few points
appear in region 2.

In order to analyse these points, and more precisely the characteristics of
their proximities, we consider the region $0.44\leq y\leq 0.56$ and $0\leq
k\leq 0.09$ with the refinement $r\geq 0.98$, as shown in Fig. \ref%
{Fig:appendix_and_double_peak}. 
\begin{figure}[tbh]
\begin{center}
\includegraphics[width=1.0\columnwidth]{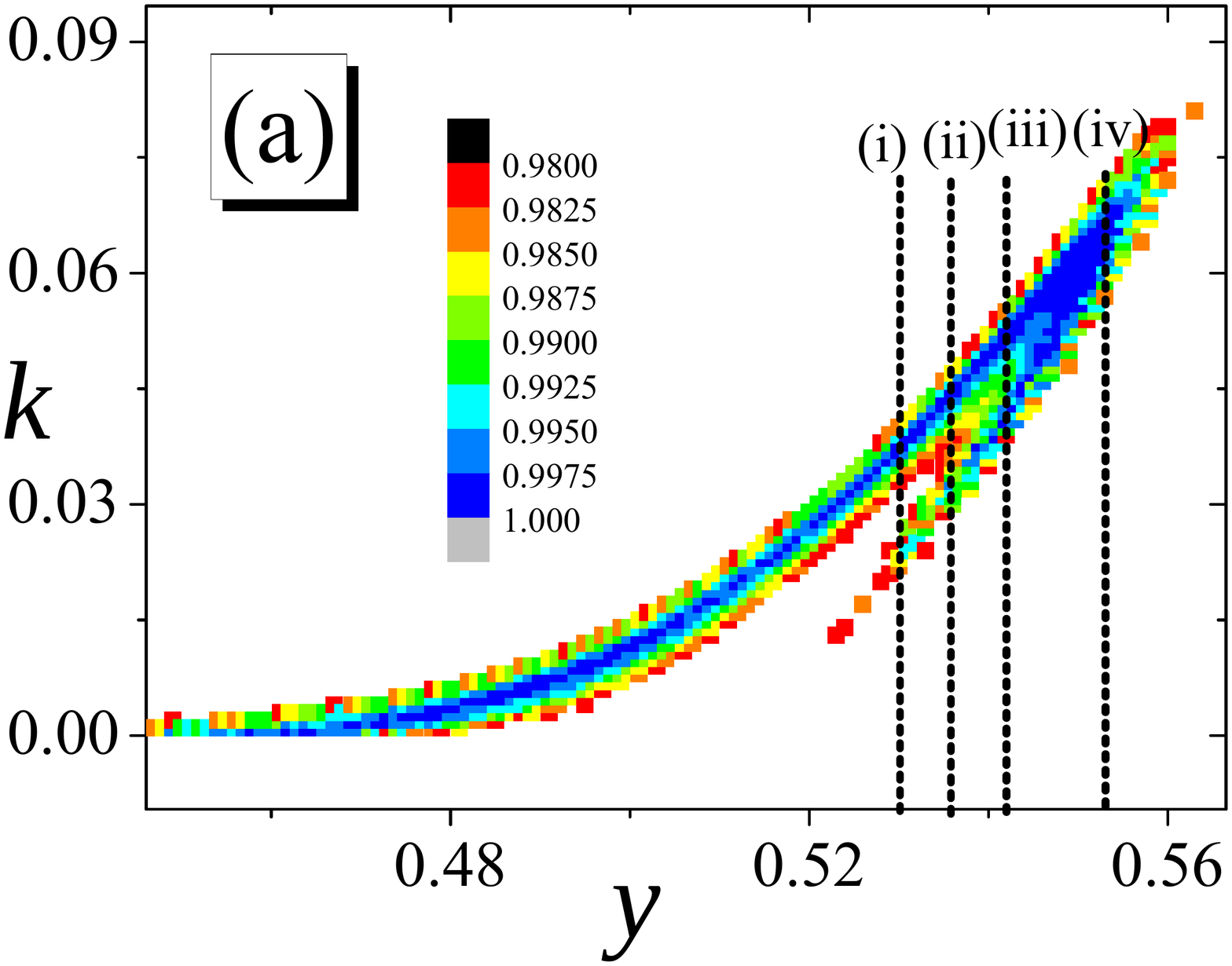} %
\includegraphics[width=1.0\columnwidth]{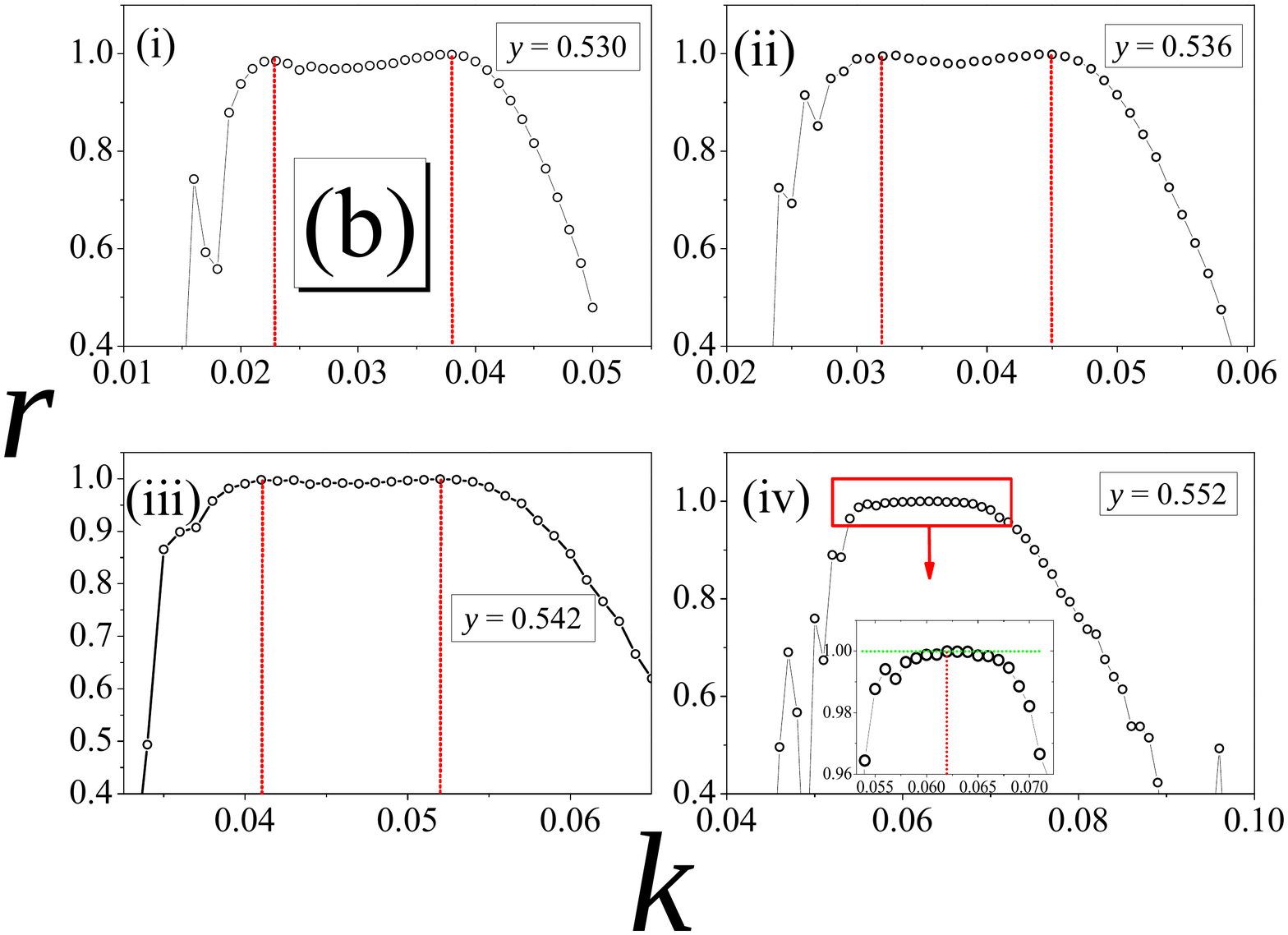}
\end{center}
\caption{Figure (a) shows a small part of the region 2 ($0.44\lesssim
y\lesssim 0.56$ and $k\lesssim 0.09$) corresponding to the two curves that,
after the refinement, lead to a single point. The vertical dashed lines
correspond to different values of $y$: (i) $y=0.530$, (ii) $y=0.536$, (iii) $%
y=0.542$, and (iv) $y=0.552,$ where the coefficient of determination $r$ is
calculated as function of $k$. Figure (b) shows $r\times k$ for these values
of $y$. We can observe that the double peaks observed in the plots (i),
(ii), and (iii) culminates in an single peak shown in the plot (iv). The
inset in the plot (d) is a zoom in order to verify the existence of a unique
peak.}
\label{Fig:appendix_and_double_peak}
\end{figure}
As can be seen, this figure highlights the two curves presented in Fig. \ref%
{fig:refinement} (a) which lie at a give point. In order to characterize the
phenomena related to this region we choose four values of $y$: 0.530, 0.536,
0.542, and 0.552, and calculate the coefficient of determination $r$ for $%
0\leq k\leq 0.09$ with $\Delta k=0.001$. The vertical dashed lines presented
in Fig. \ref{Fig:appendix_and_double_peak} (a) indicate where our analysis
is performed. We can observe that three of these vertical lines cross both
curves which means a double peak in a plot of $r\times k$ as shown in Fig. %
\ref{Fig:appendix_and_double_peak} (b) for: (i) $y=0.530$, (ii) $y=0.536$,
and (iii) $y=0.542$. However, the vertical line for $y=0.552$ [Fig. \ref%
{Fig:appendix_and_double_peak} (a)] crosses the curve only once, i.e., there
is only a single peak in Fig. \ref{Fig:appendix_and_double_peak} (b): (iv).
The inset plot in this figure corresponds only to a zoom which indeed
verifies the existence of a unique peak. As found in previous works \cite%
{albano2001b,fernandes2016}, the two peaks are related to pseudo critical
points (here interpreted from a nonequilibrium simulations' point of view)
which characterize the discontinous phase transitions (weak first order
phase transitions). Here, it is worth to mention that in a previous work, Tom%
\'{e} and Dickman \cite{tome1993} found, for $y=0.54212$ and $k=0.0406$, an
critical adsorption rate associated with an Ising-like point. However, in
our study, we believe that this point must be related to a single peak, as
shown in the last plot of Fig. \ref{Fig:appendix_and_double_peak} (b).
According to our simulations the best candidate to this Ising-like point is
slightly after $y=0.552$ and $k=0.064$. In fact, our best value of the
coefficient of determination, $r=0.99984$ was obtained for $y=0.554$ and $%
k=0.064$.

At this point, two important questions arise. Firstly, as shown in Ref. \cite%
{tome1993}, is this point an Ising-like critical point? Secondly, is the
narrow extension of points that grow up from $k=0$ to $k=1$ nearby $y\approx
0.390$ critical? If so, what can we say about its universality class?

In the next subsection, we look into some points of the region 1, Fig. \ref%
{fig:refinement} (d), and the point corresponding to the best value of $r$
of the region 2, $y=0.554$ and $k=0.064$. The analysis is carried out by
means of nonequilibrium MC simulations.

\subsection{Critical exponents}

In this subsection, we obtain the critical exponents $\delta =\beta /\nu
_{\parallel }$ and $\theta $ for some critical points through nonequilibrium
MC simulations of the ZGB model with $CO$ desorption. All results were
obtained for the following set of parameters: $L=160$, $N_{MC}=300$, and
5000 samples. The results are averages obtained through 5 different time
evolutions and the error bars are obtained from them. To study the region 1,
we decided to consider some values of $k$, equally spaced, and look for
values of $y_{\text{best}}$ which correspond to the best value of $r_{\text{%
best}}$. The values of $k$, $y_{\text{best}}$, and $r_{\text{best}}$ are
shown in Table \ref{Table:DP}. As shown in the second row of this table, the
values of $y$ are very close to the value of the critical adsorption rate of
the original ZGB model, $y=0.388$.

\begin{table*}[tbp] \centering%
\begin{tabular}{llllllllll}
\hline\hline
$k$ & $0.1$ & $0.2$ & $0.3$ & $0.4$ & $0.5$ & $0.6$ & $0.7$ & $0.8$ & $0.9$
\\ \hline
$y_{\text{best}}$ & $0.387$ & $0.388$ & $0.387$ & $0.388$ & $0.391$ & $0.388$
& $0.391$ & $0.390$ & $0.392$ \\ 
$\delta _{\text{best}}$ & $0.4838(6)$ & $0.4527(4)$ & $0.4797(7)$ & $%
0.4587(6)$ & $0.4065(2)$ & $0.4581(4)$ & $0.4146(5)$ & $0.4218(2)$ & $%
0.3953(1)$ \\ 
$r_{\text{best}}$ & $0.999784$ & $0.999767$ & $0.999801$ & $0.999867$ & $%
0.999718$ & $0.999928$ & $0.999915$ & $0.999953$ & $0.999975$ \\ 
$\delta $ & $0.4532(3)$ & $0.4527(3)$ & $0.4546(4)$ & $0.4587(6)$ & $%
0.4587(5)$ & $0.4581(4)$ & $0.4515(4)$ & $0.4396(5)$ & $0.4153(3)$ \\ 
$r$ & $0.999770$ & $0.999767$ & $0.999650$ & $0.999867$ & $0.999541$ & $%
0.999928$ & $0.999898$ & $0.999469$ & $0.999829$ \\ 
$\theta $ & $0.24(1)$ & $0.23(2)$ & $0.23(2)$ & $0.23(1)$ & $0.22(1)$ & $%
0.21(2)$ & $0.21(1)$ & $0.21(1)$ & $0.21(1)$ \\ \hline\hline
\end{tabular}%
\caption{Critical exponents estimated for the refined points of the region 1}%
\label{Table:DP}%
\end{table*}%

In order to obtain the critical exponents, we first consider the simulations
starting with the initial condition $\rho _{0}=0$, i.e., where all sites of
the lattice are initially occupied by $CO$ molecules, and we also expect a
power law described by Eq. (\ref{eq:p2}). Figure \ref%
{Fig:directed_percolation} (a) ilustratres the time evolution of $\rho (t)$
in $\log \times \log $ scale for one particular case: $k=0.1$ and $y=0.387$.
We can observe that after an initial transient, $\rho (t)$ decays (the inset
plot shows the corresponding linear fit in $\log \times \log $ scale). The
error bars taken over 5 different diferent seeds are indeed very small. The
values obtained for $\delta \ (\delta _{\text{best}})$ corresponding to the
best $y=y_{\text{best}}$ are shown in the third row of Table \ref{Table:DP}.
The values are close to $\delta =0.451$ \cite{voigt1997}. However, the
results present a certain variation and do not permit us to assert that the
points found in region 1 belong to the DP universality class. The values of
coefficient of determination for these points are presented in the fourth
row of Table \ref{Table:DP}.

Since the values of $y$ (second row) are very close to $y=0.388$ (in fact,
this value appears in four of the nine values presented in Table \ref%
{Table:DP}) we also carried out simulations by considering the same values
of $k$ but kept $y=0.388$. The coefficient of determination is still very
close to 1, as shown in the sixth row of Table \ref{Table:DP}, and the
results for the exponent $\delta $ are surprisingly. Now we have $\delta $
even more closer to the value of the DP universality class (see the row 5 of
the table).

\begin{figure}[tbh]
\begin{center}
\includegraphics[width=1.0\columnwidth]{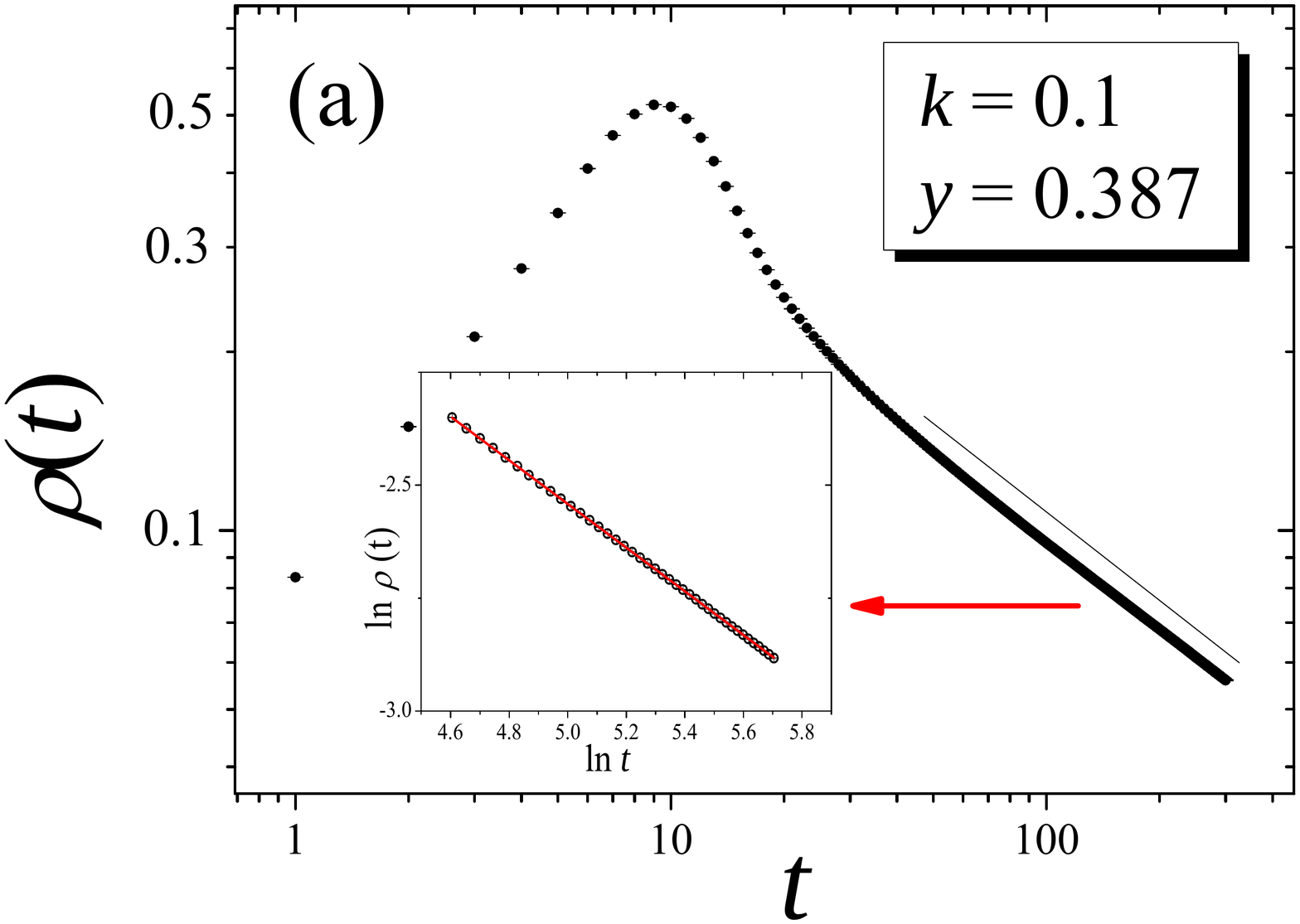} %
\includegraphics[width=1.0%
\columnwidth]{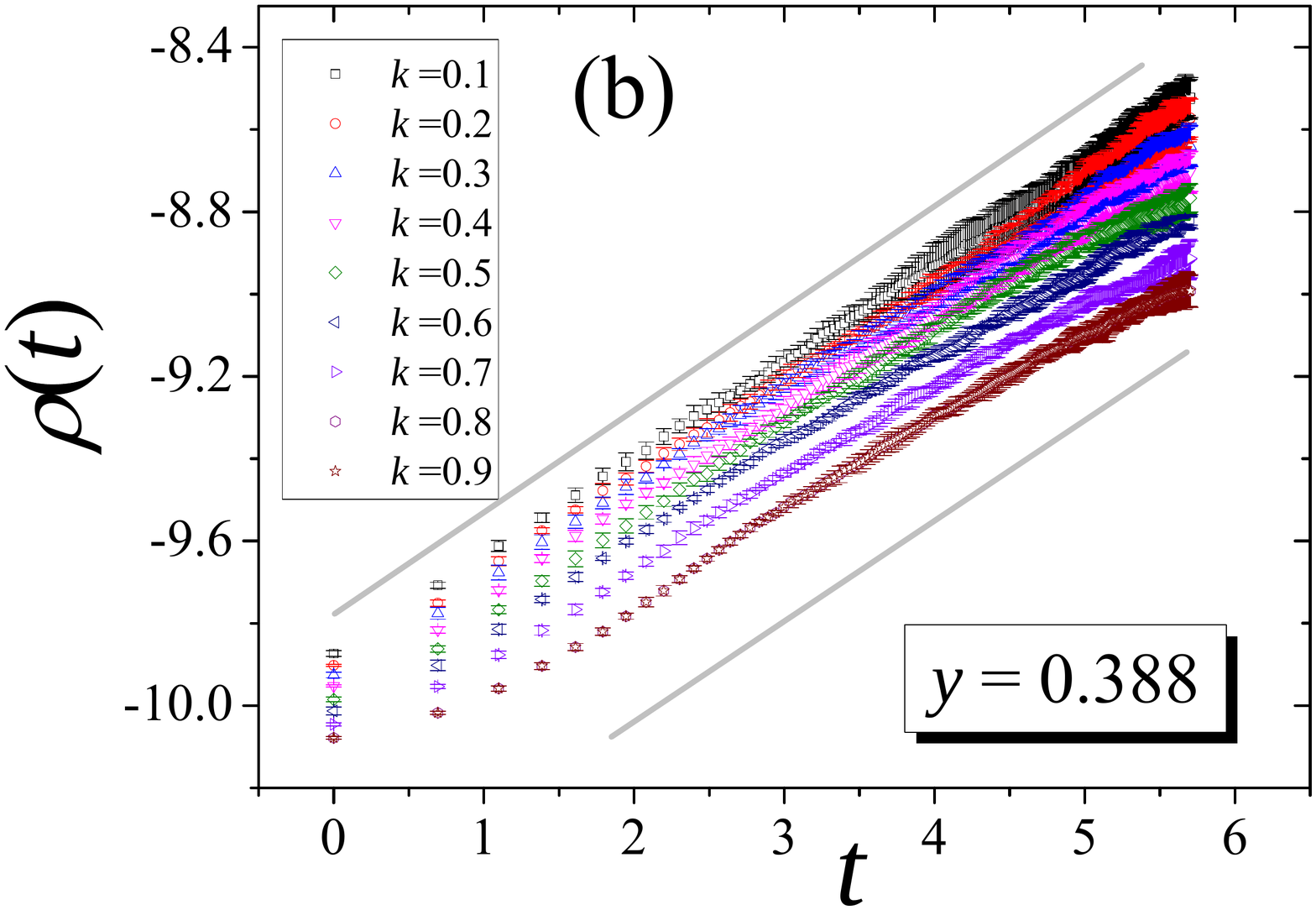}
\end{center}
\caption{Plot (a) shows the time evolution of $\protect\rho (t)$ in $\log
\times \log $ scale for one particular case: $y=0.387$ and $k=0.1$ when all
sites are inittialy filled with $CO$ molecules. We can observe that after a
transient, the system decays according to the power law given by Eq. (%
\protect\ref{eq:p1}) (the inset plot highlights the robustness of this power
law). Plot (b) shows the initial growing of $\protect\rho (t)$ when the
initial lattice is fully ocuppied by $O$ atoms except for a vacant site
located at the center of the lattice. We show the power laws for $k=0.1$, $%
0.2$, ..., $0.9$ and $y=0.388$.}
\label{Fig:directed_percolation}
\end{figure}

Thus, our results suggest that the points for the region 1 are critical ones
and may belong to the DP universality class, but further studies must be
performed in order to confirm (or not) this assertion. However, we have two
more results to present in this work and which reinforces our previous
estimates. First, we show the results related to the exponent $\theta $.
Here, it is important to note that the initial transient where $\rho (t)$
increases before decaying as $\rho (t)\sim t^{-\delta }$, does not
correspond to the critical initial slip where the exponent $\theta $ is
usually obtained. In fact, this exponent is not expected for that initial
condition. In addition, this initial transient lasts only for few steps
(less than 10 MC steps). So we performed simulations for those same points
presented in Table \ref{Table:DP} with the initial condition where all sites
are ocuppied by $O$ atoms but a single site that remains vacant at the
center of the lattice, i.e., $\rho _{0}=1/L^{2}$. The time evolution of $%
\rho (t)$ (given by Eq. (\ref{eq:p2})) is shown in Fig.\ref%
{Fig:directed_percolation} (b). The gray lines introducted in this figure
work as directions to observe that power laws are approximately parallel
lines in $\log \times \log $ scale. The seventh row of Table \ref{Table:DP}
presents the exponents $\theta $ for all considered points. These values are
very close to $\theta =0.230$ found for the two-dimensional DP model \cite%
{voigt1997}. The error bars are bigger for this initial condition as can
also be observed in Fig. \ref{Fig:directed_percolation} (b) generating
uncertainties in the second decimal digit.

\begin{figure}[tbh]
\begin{center}
\includegraphics[width=1.0\columnwidth]{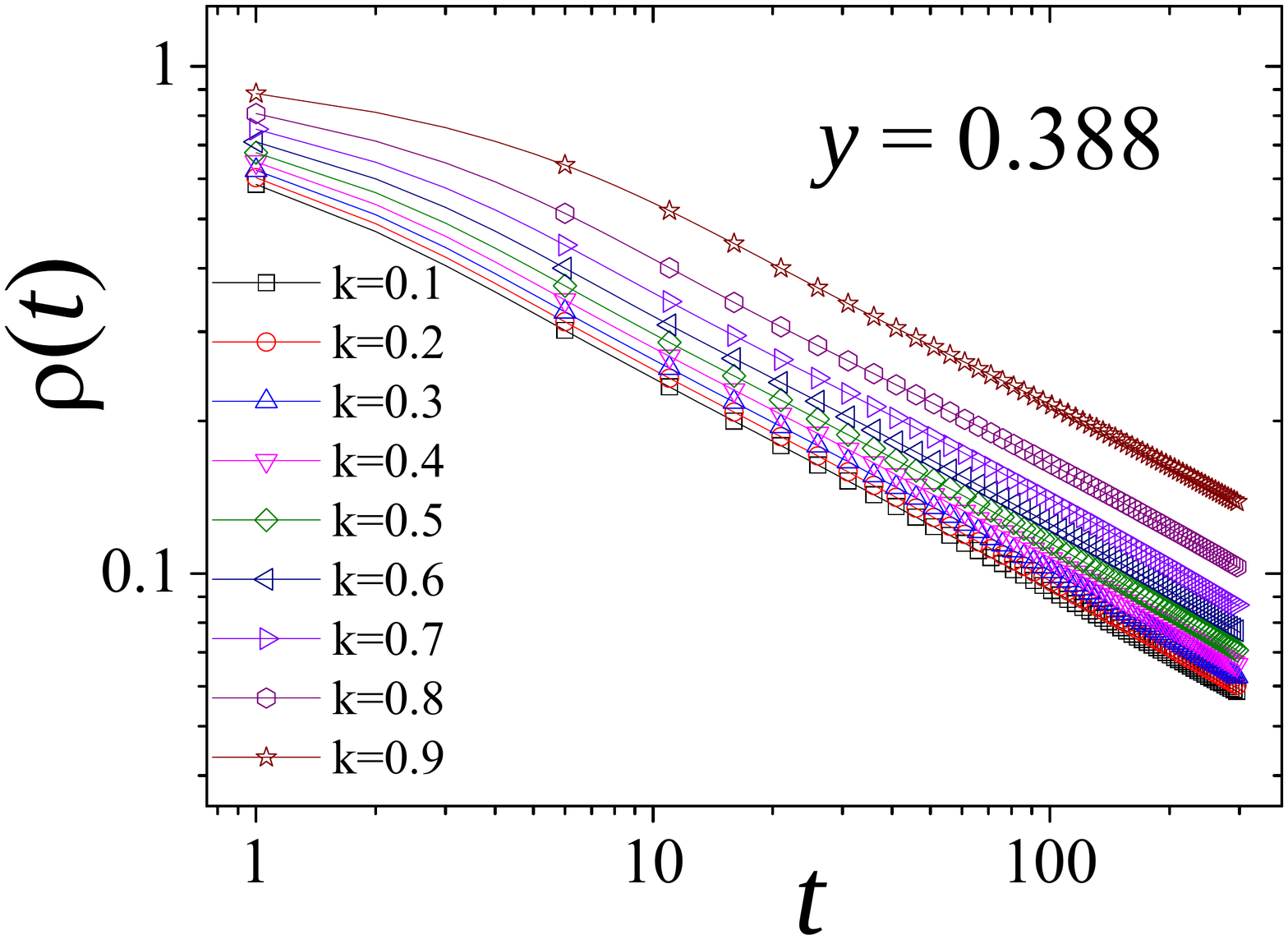}
\end{center}
\caption{Time evolution of $\protect\rho (t)$ for $y=0.388$ when we start
with all sites vacant ($\protect\rho _{0}=1$) for $k=0.1$, $0.2$, ..., $0.9$.
}
\label{Fig:time_evolution_starting_from_vacant_DP}
\end{figure}

Secondly, it is also possible to study the behavior of $\rho (t)\sim
t^{-\delta }$ when all initial sites are vacant, since we expect this
behavior specifically for this initial condition when considering the
density of vacant site as the order parameter of the model. We verifed that,
in this case, the exponent $\delta $ corresponds exactly to what is expected
when $\rho _{0}=1$, giving $\delta \approx 0.45$ for all values of $k$ as
can be observed in Fig. \ref{Fig:time_evolution_starting_from_vacant_DP}
which present curves with aproximately the same slope (parallel lines).

\begin{figure}[tbh]
\begin{center}
\includegraphics[width=1.0\columnwidth]{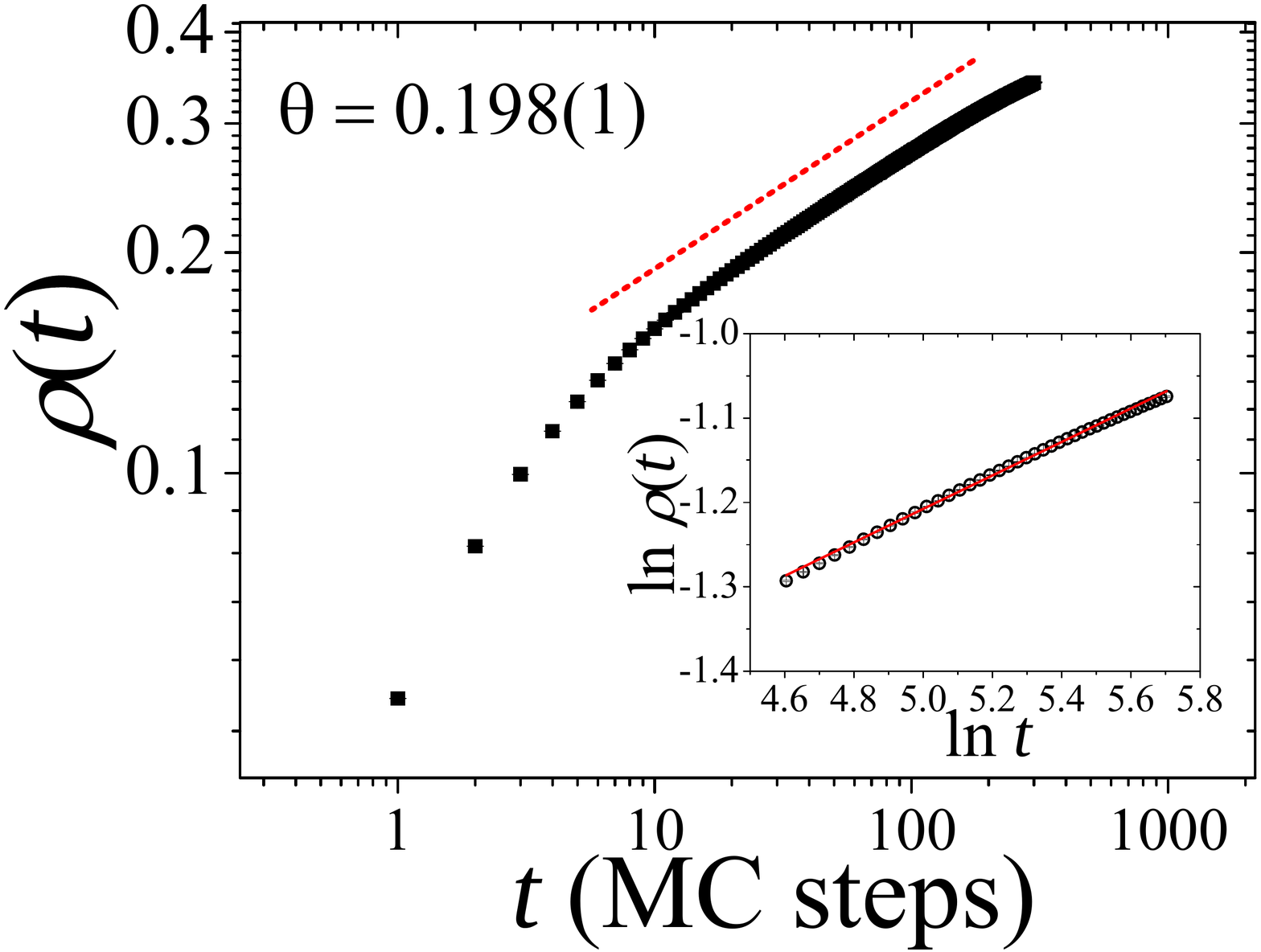}
\end{center}
\caption{Time evolution of density of vacant sites, $\protect\rho (t)\times t
$, for $y=0.554$ and $k=0.064$ when we start the simulation with all sites
ocuppied by $CO$ molecules. The initial slip corresponds to the exponent $%
\protect\theta $.}
\label{Fig:Ising_model}
\end{figure}

Since we unveiled the region 1, let us briefly explore the region 2 which,
according to the previous subsection, culminates in a point corresponding to
a unique peak for the coefficient of determination, after a succession of
double peaks (pseudo critical points). For this point ($y=0.554$ and $k=0.064
$), we consider the initial lattice being fully occupied with $CO$
molecules. Figure \ref{Fig:Ising_model} shows the critical initial slip for
the density of vacant sites as function of $t$, in $\log \times \log $
scale. 
The slope of this curve gives 
\begin{equation}
\theta =0.198(1)  \label{Eq:estimates}
\end{equation}%
and is in fair agreement with the results found for the Ising model, $\theta
\approx 0.193$ \cite{zheng1998}. Our estimate identifies this critical point
as belonging to the Ising universality class, exactly as predicted in Ref.%
\cite{tome1993} for the static exponent $\nu $ for a point slithly before
ours, $y=0.54212$ and $k=0.0406$.

In this work, we did not perform a complete study of the critical exponents
of this point since this was not our main purpose and, in addition, the
topic deserves a better compreension about the exact relationship between
magnetic models and absorbing state models. We also verified that the choise
of both order parameter and the initial condition seems to be very important
to estimate the critical exponents. So, these topics will be addressed in a
further work.

\section{Conclusions}

\label{sec:conclusions}

In this paper, we studied the ZGB model with desorption of $CO$ molecules
through nonequilibrium Monte Carlo simulations and showed for the first time
that it can belong to the DP and Ising universality classes, depending on
the values of the $CO$ adsorption rate, $y$, and of the $CO$ desorption
rate, $k$. We presented the diagram $k\times y$ obtained through the
optimization of the coefficient of determination and showed that the region
belonging to the DP universality class is composed by a line of critical
points extending from $0\leq k\leq 1$ and $y\simeq 0.388$. The other region
possesses two pseudo critical lines that probably intercept each other at
the Ising-like critical point, $y=0.554$ and $k=0.064$.

\section*{Acknowledgments}

This research work was in part supported financially by CNPq (National
Council for Scientific and Technological Development). This work was partly
developed using the resources of HPCC - High Performance Computer Center -
Jata\'{\i}. R. da Silva would like to thank L. G. Brunnet
(IF-UFRGS) for kindly providing resources from Clustered Computing Center
Ada Lovelace for the partial development of this work.

\end{document}